# TO DEEPFAKE OR NOT TO DEEPFAKE: HIGHER EDUCATION STAKEHOLDERS' PERCEPTIONS AND INTENTIONS TOWARDS SYNTHETIC MEDIA

Jasper Roe [1*], Mike Perkins [2], Klaire Somoray [3], Dan Miller [3], Leon Furze [4],

[1] Durham University, United Kingdom
[2] British University Vietnam, Vietnam
[3] James Cook University, Australia
[4] Deakin University, Australia

[*] Corresponding Author: jasper.j.roe@durham.ac.uk

February 2025

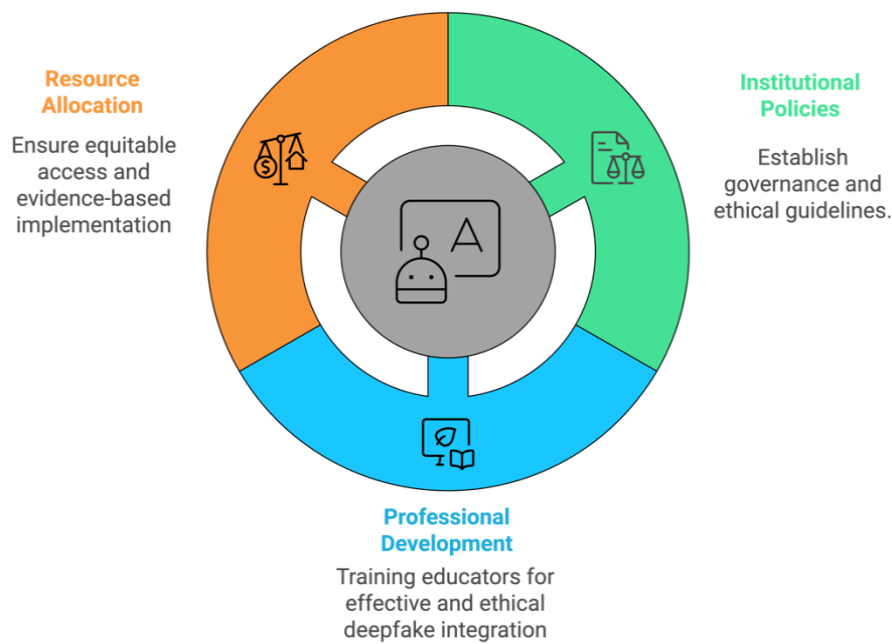

*Deepfake Adoption Framework*

# Abstract

Advances in deepfake technologies, which use generative artificial intelligence (GenAI) to mimic a person's likeness or voice, have led to growing interest in their use in educational contexts. However, little is known about how key stakeholders perceive and intend to use these tools. This study investigated higher education stakeholder perceptions and intentions regarding deepfakes through the lens of the Unified Theory of Acceptance and Use of Technology 2 (UTAUT2).

Using a mixed-methods approach combining survey data ($n$=174) with qualitative interviews, we found that academic stakeholders demonstrated a relatively low intention to adopt these technologies ($M$=41.55, $SD$=34.14) and held complex views about their implementation. Quantitative analysis revealed adoption intentions were primarily driven by hedonic motivation, with a gender-specific interaction in price-value evaluations. Qualitative findings highlighted potential benefits of enhanced student engagement, improved accessibility, and reduced workload in content creation, but concerns regarding the exploitation of academic labour, institutional cost-cutting leading to automation, degradation of relationships in education, and broader societal impacts.

Based on these findings, we propose a framework for implementing deepfake technologies in higher education that addresses institutional policies, professional development, and equitable resource allocation to thoughtfully integrate AI while maintaining academic integrity and professional autonomy.

# Introduction

Deepfakes are media outputs created through Artificial Intelligence (AI) applications which mimic a person's likeness or voice. The term 'deepfake' was coined by a user on the social media website Reddit and is etymologically derived from a combination of 'deep learning' and 'fake' (Kietzmann et al., 2020). Although some use the term 'synthetic media' to describe such outputs (Pawelec, 2024), we distinguish between these terms on the basis that a deepfake emulates a *real* person (living or dead), whereas synthetic media may depict a person who has never existed in reality.

Deepfake technologies are becoming increasingly available, accurate, and difficult to distinguish from reality (Roe et al., 2024) and are associated with malicious use cases, as they can depict people saying and doing things that they did not actually say or do (Fallis, 2021). Deepfakes have been involved in the creation of non-consensual explicit materials (Delfino, 2019) and political disinformation, fraud, and market manipulation (Langguth et al., 2021); thus, the term has a strong negative connotation. Despite the myriad risks of deepfake technology, there are potentially beneficial applications being discussed in the field of education (Danry et al., 2022; Roe, Perkins, & Furze, 2024), including the ability of learners to converse with historical figures and receive instantaneous translations (Gaur & Arora, 2022), or give a voice to those who are unable to speak (de Ruiter, 2021). Empirical studies have suggested that students may find value in using AI-generated avatars in online learning content (Vallis et al., 2024). However, in their review of the deepfake literature, Roe et al. (2024) identified a significant research gap: despite the technology's growing prominence, few empirical studies have specifically investigated the application of deepfakes and synthetic media in higher education settings.

Given the novelty of deepfake technology in education, understanding the factors that influence its acceptance is crucial. The Unified Theory of Acceptance and Use of Technology 2 (UTAUT2) provides a comprehensive framework for examining technology adoption in consumer contexts, making it particularly suitable for studying higher education stakeholders' acceptance of deepfake technology. In this study, we investigated higher education employees (educators, researchers, and administrators and leaders) perceptions and intentions regarding the use of deepfake technologies in higher education using the UTAUT2 framework to determine which factors may significantly influence the intention to use deepfakes. Second, we elicited and explored qualitative data on stakeholder views surrounding the opportunities and problems posed by deepfakes in higher education teaching and learning settings.

Our results enhance the understanding of the role emerging technologies (such as deepfakes) could play in higher education teaching and learning, drawing on a cross-functional, international sample. This offers a basis for further research to construct HE policy development and ethical guidelines.

# Literature review

**Deepfake Research in Education**

Evidence suggests little public awareness of deepfakes. In the German context, a large-scale study of Internet users (*n*=1,421) showed that most respondents had little knowledge of deepfakes (Bitton et al., 2024). However, academic research in this area is still developing. Godulla et al. (2021) identified through a systematic review that the majority of research in deepfake technologies centres on computer science, politics, and law. In the field of education, Murillo-Ligorred et al. (2023) explored 100 postgraduate students' awareness of deepfake imagery, noting that older (above 20 years) students felt more confident identifying deepfakes than younger (below 20 years) students. Erduran (2024) posited that deepfakes may be used to improve education by developing simulations for learning. Doss et al. (2023) in a large-scale study across educators, students, and the general population found that between 27-50% of respondents could not distinguish between authentic and deepfake videos, with adults and educators showing lower detection accuracy than students. No study has yet explored the topic of educators using their own deepfakes to create educational content, although extant technologies which create 'digital avatars' of individuals, such as HeyGen, are being marketed to educators and higher education institutions (HEIs) (Heygen, 2025). Roe et al.'s (2024) scoping review develops a research agenda for exploring this technology in higher education, including suggesting exploring the attitudes of higher education stakeholders.

**Theoretical Basis: From TAM to UTAUT2**

Understanding the factors that influence technology adoption is vital for its successful implementation. The Technology Acceptance Model (TAM; Davis, 1989) is a widely applied framework for technology adoption and has been extensively utilised to examine the adoption of emerging educational technologies such as virtual learning technologies and mobile learning (Granić & Marangunić, 2019). Initially conceptualised from the Theory of Reasoned Action (Fishbein & Azjen, 1975), TAM contends that there are two primary factors influencing technology adoption: perceived usefulness (the degree to which the technology enhances job performance) and perceived ease of use (how easy it is to learn and use the system) (Davis, 1989). Perceived usefulness and perceived ease of use are hypothesised to influence attitudes towards technology adoption, which in turn impacts the behavioural intention to use the technology. Therefore, if educators find deepfake technology cumbersome and difficult to use, they may resist using it, regardless of its potential benefits. Indeed, Scherer et al.'s (2019) meta-analysis of TAM in educational technology adoption found that perceived usefulness has a significant direct effect on behavioural intention ($b = 0.366$), and the overall model explained 40.1% and 31.1% of the variance found in intention and technology use, respectively.

While TAM's parsimony has contributed to its widespread adoption across various contexts, other studies have found critical limitations in this framework (Ajibade, 2018). For example, TAM's focus on individual perceptions neglects the complex social and organisational dynamics that influence technology adoption, particularly in organisations where peer influence and institutional support play crucial roles. These limitations became evident as researchers attempted to apply the TAM to increasingly sophisticated technologies. For instance, studies examining the adoption of AI as a pedagogical tool found that while perceived usefulness and ease of use were significant factors, other factors were required to explain the variance in behavioural intention (Al Darayseh, 2023; Kavitha & Joshith, 2025). Kavitha and Joshith (2025) found that educators' "inherent openness" (p. 17) to technology as well as their general sense of self-efficacy with digital technologies impacted their intentions to use AI tools, while Al Darayseh (2023) reported that perceived ease of use, expected benefits, and attitudes accounted for significant variance in science teachers' AI adoption intentions.

In response to these limitations, researchers have developed and adapted the original model. Additional factors, such as social influence and cognitive factors, have been added to the model (TAM2; Venkatesh & Davis, (2000). A more comprehensive framework was later proposed: the Unified Theory of Acceptance and Use of Technology (UTAUT) (Venkatesh et al., 2003). Within this framework, performance expectancy, effort expectancy, social influence, and facilitating conditions were deemed direct determinants of user acceptance and usage behaviour. Moderating variables such as gender, age, experience, and voluntariness of use were also added to the framework. Similar to perceived usefulness, performance expectancy is defined as the degree to which an individual believes that using the technology will improve their job performance. Similar to the perceived ease of use, effort expectancy is defined as the degree of ease associated with the use of new technology. Social influence refers to the degree to which an individual perceives that others believe he or she should use the new system, and facilitating conditions refer to the degree to which an individual believes that an organisational and technical infrastructure exists to support the use of the system. This model was then refined into UTAUT2 (Venkatesh et al., 2012), adding price value, habit, and hedonic motivation as predictor variables.

UTAUT2's application to AI-enabled educational technologies can provide insights into the further adoption of emerging technologies such as deepfakes as educational tools. For instance, Strzelecki et al. (2024) applied UTAUT2 to examine the use of ChatGPT in academic work. The findings demonstrate that the model explains 74.4% of the variance in the behavioural intention to use ChatGPT, with habit, performance expectancy, and hedonic motivation as the strongest predictors of behavioural intention. However, other studies applying UTAUT2 to the adoption of ChatGPT by educators have also revealed some limitations of the framework. For example, Eldakar et al. (2025) study of Egyptian academics demonstrated that the perceived ethics of generative AI strongly and significantly predicted the intention to use it in scientific research. These findings highlight both the utility and limitations of the UTAUT2 in understanding the adoption of emerging educational technologies. The unique characteristics of deepfake technology, including its potential for misuse, ethical implications, and nascent state of development, suggest the need to explore other factors to complement this model.

## Current study

Given the nascent state of deepfake technology in education and the absence of empirical research examining its adoption (Roe et al., 2024), this study adopted an exploratory approach to examine higher education stakeholders' adoption of deepfake technology. We focus not only on educators but also on all those working in HE who are involved in core university activities. The rationale behind such an approach is that higher education learning design and technology adoption is not wholly decided by the instructor; there can be varying pressures, reasons, or preferences for adopting new technologies in higher education teaching and learning. As this is an exploratory study, it is important to gain access to a wide range of perspectives on the topic. Furthermore, while the UTAUT2 framework provides a theoretical foundation for understanding technology acceptance, the unique characteristics of deepfake technology suggest the need for a more open-ended investigation. Therefore, our study aimed to understand the factors influencing educators' adoption intentions regarding deepfake technology in their teaching practice using the UTAUT2 framework, supplemented by open-ended questions about the perceived challenges and benefits of deepfake technology in educational settings. Instead of testing predetermined hypotheses, we posed the following research questions.

1. What factors influence higher education stakeholders' (educators, researchers, administrators, and leaders) perceptions and intentions regarding the use of deepfake technologies in higher education, as framed by the UTAUT2 (Unified Theory of Acceptance and Use of Technology 2) model?
2. What potential benefits or risks do these stakeholders perceive in the use of these technologies in higher education?

## Methods

**Procedure**

The current study used an online cross-sectional design with convenience sampling to recruit the participants. Participants were recruited via mailing lists of higher educational institutions and professional network platform posts (e.g. LinkedIn). All participation was voluntary, and the participants received no compensation for their involvement in the study. Informed consent was obtained to access the survey, and as an anonymous survey, this study was granted research ethics exemption from an institutional review board, and the [research questions were pre-registered on Research Box](#). To participate, respondents had to confirm that they were above 18 years of age and worked in a research, teaching, or senior leadership role in a higher education context. The survey was administered using the Qualtrics platform.

**Participants**

A total of 258 participants completed a part of the survey. Of the 258, 16 did not complete any items beyond the consent question, and 49 completed demographic items only. This resulted in a final sample size of *n* = 192. Of these participants, 62.5% self-identified as female and 33.9% as male. The rest identified as non-binary (0.5%) or did not report their gender (3.1%). The mean age of the sample was 46.7 years ($SD$ = 10.3). Most participants reported being in a teaching-focused (*n* = 80, 41.7%) or a balanced teaching and research role (*n* = 76, 39.6%). Fewer participants held research-focused (*n* = 17, 8.9%) or senior management positions (*n* = 17, 8.9%).

**Data collection and measures**

The online survey included questions relating to gender, age, and academic role, information about deepfakes with a brief video as an example, and examples of how deepfakes may be used in education (e.g. using an academic's likeness to deliver video lectures). Both quantitative and qualitative data were collected because of the novelty of the topic, measuring specific constructs related to technology acceptance using the UTAUT2 framework (quantitative) and open-ended questions on educators' perceived concerns and benefits of deepfake technology for educational purposes (qualitative).

The seven UTAUT2 constructs used were performance expectancy, effort expectancy, social influence, facilitating conditions, hedonic motivation, price value, habit, and behavioural intention. All items used a 100-point slider scale ranging from '0 - Strongly Disagree' to '100 - Strongly Agree'. Example items included: 'I believe deepfakes could be used effectively to create engaging educational content' (performance expectancy); 'I typically make use of new/innovative technology in my teaching' (habit). The UTAUT2 constructs were developed by the researchers using validated scales in the literature (e.g. Davis, 1989), with modifications made to specifically address deepfake technology in educational contexts. The items used in this study are listed in Appendix 1. The qualitative component consisted of two open-ended questions that examined the perceived benefits and concerns regarding the use of deepfake technology in education. The measures used and data for this study are available at [Research Box](#).

## Quantitative Results and Analysis

**Quantitative Analysis**

*Data Cleaning and Preliminary Analysis*

Due to the very small number of non-binary participants ($n = 1$), this group was excluded from any statistical analysis involving gender. Furthermore, participants who did not respond to the primary outcome variable (behavioural intention to use deepfakes in one's teaching practice) were excluded ($n = 17$), leaving a final sample of 174 for quantitative analysis.

The variables were assessed for univariate outliers using boxplots. A small number of outlying datapoints ($n = 3$) were detected at the low end of the distribution for the habit variable. Given that none of these outliers were extreme, the data points were not adjusted. No other variables displayed univariate outlier data points.

Zero-order correlations were generated between the scale-level variables. These are reported in Table 1 along with the descriptive statistics. On average, participants rated their intention to use deepfakes as part of their teaching practice at 41.55 ($SD = 34.14$) on a scale of 0–100, suggesting a lower intention to adopt this technology. The observed range for this variable was 0 to 100, indicating a diversity of opinions regarding the use of deepfakes among the sample. Except for age and gender, all predictors were significantly correlated with the behavioural intention to use deepfakes. The largest association was observed between hedonic motivation and behavioural intention ($r = .83$), suggesting that people who thought deepfakes would be enjoyable or interesting to use were more likely to use them for teaching purposes. The predictor variables also tended to be significantly and positively correlated with each other.

**Table 1: Descriptive Statistics and Zero-Order Correlations for Study Variables**

|  | Intentions | Performance Expectation | Effort Expectation | Social Influence | Hedonic Motivation | Price Value | Habit | Age | Gender |
|---|---|---|---|---|---|---|---|---|---|
| **Mean (*SD*)** | 41.55 (34.14) | 33.68 (25.07) | 44.54 (31.83) | 32.65 (26.34) | 42.07 (34.40) | 35.75 (30.76) | 71.84 (23.84) | 46.72 (10.34) | NA |
| **Observed Range** | 0.00 – 100.00 | 0.00 – 100.00 | 0.00 – 100.00 | 0.00 – 100.00 | 0.00 – 100.00 | 0.00 – 100.00 | 0.00 – 100.00 | 24.00 – 74.00 | NA |
| **Intentions** | — | | | | | | | | |
| **Performance Expectation** | .59*** | — | | | | | | | |
| **Effort Expectation** | .42*** | .44*** | — | | | | | | |
| **Social Influence** | .56*** | .66*** | .30*** | — | | | | | |
| **Hedonic Motives** | .83*** | .55*** | .53*** | .50*** | — | | | | |
| **Price Value** | .40*** | .26*** | .09 | .30*** | .33*** | — | | | |
| **Habit** | .28*** | .17* | .44*** | .14 | .33*** | –.01 | — | | |
| **Age** | –.14 | –.17* | –.10 | –.09 | –.07 | –.13 | –.05 | — | |
| **Gender (female†; male)** | .03 | .12 | .11 | .04 | .08 | –.01 | –.12 | –.19* | — |

*Note*. *df* = 158-171
* *p* < .01; ** *p* < .01; *** *p* < .001. † indicates the reference category.

*Statistical findings*

A linear regression model predicting the behavioural intention to use deepfakes in one's teaching practice was specified. Mirroring the UTAUT2 framework, eight one-way predictors (performance expectancy, effort expectancy, social influence, hedonic motivation, price value, habit, age, and gender) and six interaction terms (age × hedonic motivation, age × price value, age × habit, gender × hedonic motivation, gender × price value, and gender × habit) were entered into the model.

For this initial model, a scatterplot of residual versus predicted values indicated normality, linearity, and homoscedasticity of model residuals. Multicollinearity was tested with reference to variance inflation factor (VIF) values. Some VIFs were large; however, for models with interaction terms, high values on measures of collinearity are possible (Cohen et al., 2003) and should not necessarily be considered problematic (Hayes, 2018). Cook's distance values were small (ranging from 0.00 to 0.18) indicating a lack of influential outliers.

The initial model accounted for a statistically significant portion of the variance in behavioural intention, $F(14, 138) = 30.08$, $p < .001$, $R^2 = .75$, and adjusted $R^2 = .73$. Following Hayes's (2018, p. 231) recommendations, non-significant interaction terms were iteratively dropped from the model (starting with the interaction term with the largest associated *p* value) to facilitate the interpretation of coefficient values. This process left a finalised model with eight one-way predictors and one interaction term (gender × price value). For the finalised model, a scatterplot of residual versus predicted values again suggested normality, linearity, and homoscedasticity of residuals. VIFs were all in the acceptable range (ranging from 1.08 to 2.03), except for those associated with the interaction term (gender = 5.40; price value = 1.81; gender × price value = 5.83). Cook's distance values were again small (ranging from 0.00 to 0.18).

This finalised model accounted for a statistically significant portion of the variance in behavioural intentions, $F(9, 143) = 46.52$, $p < .001$, $R^2 = .75$, and adjusted $R^2 = .73$. The coefficient values for the individual predictors are presented in Table 2. As can be seen, only hedonic motivation and gender × price value were significant predictors in the final model. The significant interaction effect was probed using Hayes' (2018) PROCESS macro (v. 4.2). Specifically, the conditional effect of price value on behavioural intention was assessed at both levels of gender (while the other variables were included as covariates). This probing revealed that price value was positively related to behavioural intention among men, $\theta_{X \to Y} | (W = male) = 0.27$, 95% CI [0.43, 0.11], $p = .002$, but not among women, $\theta_{X \to Y} | (W = female) = 0.06$, 95% CI [0.18, 0.07], $p = .352$.

**Table 2: Final Model Predicting Behavioural Intention to Use Deepfakes**

| Predictor | B | SE | t | p | β | $sr^2$ |
|---|---|---|---|---|---|---|
| Performance Expectancy | 0.15 | 0.09 | 1.74 | .085 | .10 | 0.01 |
| Effort Expectancy | −0.07 | 0.06 | −1.22 | .225 | −.06 | <.01 |
| Social Influence | 0.14 | 0.07 | 1.92 | .057 | .11 | 0.01 |
| Hedonic Motivation | 0.69 | 0.06 | 11.70 | <.001 | .70 | 0.24 |
| Price Value | 0.06 | 0.06 | −0.93 | .352 | −.05 | <.01 |
| Habit | 0.06 | 0.07 | 0.82 | .416 | .04 | <.01 |
| Gender (female†; male) | 10.60 | 7.26 | 1.46 | .146 | −.08 | 0.01 |
| Age | −0.22 | 0.14 | −1.54 | .126 | −.07 | <.01 |
| Gender × Price Value | 0.21 | 0.10 | −2.08 | .040 | −.19 | 0.01 |

† indicates the reference category.

*Quantitative Analysis Summary*

Quantitative analysis revealed two key patterns in educators' intentions to adopt deepfake technology. First, the dominant role of hedonic motivation suggests that adoption decisions may be primarily driven by the engaging and novel aspects of the technology rather than its practical utility, given that hedonic motivation emerged as the strongest predictor ($\beta = .70$), while accounting for other variables typically important in technology adoption, such as performance expectancy and effort expectancy. Second, the gender-specific interaction with price value considerations showed a pattern of male educators' adoption intentions being significantly influenced by cost-benefit evaluations, while female educators' intentions were not. This suggests that institutional approaches to implementation may need to consider gender-specific strategies when addressing resource and value considerations for the same.

While the regression model identified only hedonic motivation and price value as significant predictors of intention to adopt deepfake technology, this does not necessarily indicate that other UTAUT2 constructs are unimportant in understanding adoption decisions. Indeed, the preliminary analysis revealed moderate zero-order correlations between most UTAUT2 constructs and adoption intention. The non-significance of other predictors in the regression model likely stems from overlapping variance in explaining intention, with few constructs having explanatory power over and above the others. This interpretation is supported by the strong intercorrelations between the UTAUT2 constructs, as shown in the zero-order correlations in Table 2. Moreover, the squared semipartial correlation for hedonic motivation (see $sr^2$ column in Table 2) indicates that even without this predictor, the remaining variables would explain 51% of the variance in the intention to integrate deepfakes (model $R^2 - sr^2$ for hedonic motivation = .51).

## Qualitative Results and Analysis

Thematic Analysis (TA) has a track record of being used successfully in educational research (Ain et al., 2019; Ammigan et al., 2023). However, TA is poorly demarcated (Braun & Clarke, 2006) and requires careful explanation to demonstrate its rigor and clarify the approach used. In our analytical method, we undertook reflexive (formerly known as organic) TA, in which themes did not pre-exist the analysis but were actively and inductively constructed through the researcher's reflexive engagement (Braun & Clarke, 2016, 2019). This represents a constructivist ontological position and an interpretivist epistemological stance, which serves as a triangulation and counterbalance to our quantitative analysis. Our quantitative analysis serves to identify measurable relationships through a post-positivist orientation, while our qualitative analysis aimed to explore the meanings of how deepfake technology in education is socially constructed and interpreted by stakeholders.

Our qualitative response questions asked respondents to describe their perceptions of the benefits and challenges of deepfakes for higher education. The question on potential benefits received 157 responses, while the question on potential challenges received 166 responses, both lower than the overall response rate for the survey (258 responses). Following the six-step approach to TA (Braun & Clarke, 2006), we began with data familiarisation through close reading and then utilised open coding in NVivo 14 to identify repeated patterns of meaning. We used semantic rather than latent analysis, coding the data at the surface level rather than attempting to interpret their hidden meanings (Pigden & Jegede, 2020). Although demographic data were collected and used in the quantitative analysis, for this section, we opted to treat the data holistically to avoid trying to 'fit' qualitative data into quantitative logics (Braun & Clarke, 2019). Following coding, patterns of meaning were constructed into themes that were iteratively and reflexively refined. This led to the development of three themes, each encapsulating both the challenges and opportunities of deepfake technology in higher education: Learning and Engagement, Ethical Concerns, and Efficiency. Each theme contained several subthemes, as detailed in Table 3.

**Table 3: Themes and Subthemes**

| Theme | Subthemes | Description |
|---|---|---|
| **Impacts on the Educational Experience**<br><br>Refers to both opportunities and challenges that deepfake technology poses to learning and engagement matters. | Inclusion | The ability for deepfakes to assist with inclusivity in the learning process. |
| | Excitement and Engagement | The ability for deepfakes to generate greater engagement and stimulation for learners. |
| **Perceived Degradation of Education**<br><br>Describes issues relating to ethics surrounding the use of deepfakes in higher education. | Transparency and Literacy | Concerns about deepfakes not being clearly described to learners and teachers, and the risk of insufficient AI literacy to know when a deepfake is being shown. |
| | Misuse | Multiple forms of misuse, including for fraud, mis and disinformation, and exploitation of academics. |
| | Diminishing Human Value | The loss of centrality of human relationships and communication in the academy, and the potential for job loss through automation. |
| | Societal impacts | Large-scale impacts such as environmental and climate change, energy use, and loss of control of self-image. |
| **Modulating Educator Efficiency**<br><br>Refers to a perceived increase or decrease in efficiency over multiple processes in higher education using deepfakes. | Workload | Perceived reduction in workload as a result of deepfake technology applications. |
| | Cost-Benefit | Analysis of the potential costs of deepfake technology use versus perceived benefits in multiple domains. |

**Theme 1 – Impacts on the Educational Experience**

*Sub-theme: Inclusion*

One of the most compelling aspects of the responses was the sense that deepfake technology could be used to bolster inclusion in learning and teaching. Most frequently, this referred to translation, with many responses noting that deepfakes could be beneficial for non-native English speakers and appealing to a wider variety of learners. However, inclusion was also mentioned in multiple responses to help engage learners who found it difficult to engage with traditional learning materials, as follows:

> *"Deepfakes could be used to bring back historical figures to explain their achievements, he said. This would be helpful for less capable students who dislike or can not read!!"*

Furthermore, responses at times focused not only on visual deepfakes but also on the synthetic generation of voices, with the understanding that such a use could help students participate if they struggled in this regard:

> *"It could also help improve education accessibility for students with disability (deepfake voices could be tailored to speak with greater clarity)."*

*Sub-theme: Excitement and Engagement*

A pattern that occurred frequently throughout the data was the idea that deepfakes of historical figures could provide engagement, excitement, and a potential benefit to learners. This may be the result of a priming effect from the examples given in the introduction to the survey. In the examples below, the potential for 'reanimating' historical figures through deepfake technology to support and engage students is demonstrated:

> *"It would allow for educators to present information in a more interesting way for students if they utilised a celebrity or famous figure to explain a concept that requires students full attention."*

> *"The underlying technology could be used to create representations of historical figures, leading to more memorable video content. Perhaps it could also be used in conjunction with chatbots powered by LLMs, allowing students to "talk to" historical characters. Along the same lines, it could be used to make more immersive AR or VR simulation."*

**Theme 2 – Perceived Degradation of Education**

The most prominent pattern that recurred throughout the dataset was a deep concern regarding the potential for ethical misuse, spanning a range of different topics from individual rights to personal image to far-reaching societal impacts in education and beyond.

*Sub-theme: Transparency and Literacy*

Transparency and literacy were combined into a separate subtheme, as these concepts often overlapped. The first common ethical concern was that the use of deepfake technology in higher education would set a precedent for embracing deepfakes, which could subsequently damage students' AI literacy levels. The following extract demonstrates this concern:

*"Deep fakes have the potential to confuse with reality which could be long lasting eg a lecture by a Nelson Mandela Deepfake could be perceived as a reality, if this is later used by the student as evidence."*

The second extract evidences the concern regarding the normalisation of deepfakes in an educational setting:

*"Using it in education services would normalise the use of deepfakes more broadly. Which would potentially encourage misuse of deepfakes."*

### *Sub-theme: Diminishing Human Value*

Commonly, ethical concerns are related to the replacement of academics with non-human technologies, and subsequently diminishing the importance of a human connection in education. This aligns with current concerns in the educational literature suggesting that AI tools may negatively impact human agency in education (Roe & Perkins, 2024). This theme often recurred in terms of devaluing the job of an academic and creating an automated higher-education offering, resulting in job losses through automation and a lower-quality educational experience for students, as seen below:

*"Exploitation of the academic to create the original content and then use of deep fakes to continue teaching the content, i.e. putting the academic out of a job"*

*"Just about every bloody thing! Essentially, to save money and time, deepfakes would devalue the personalised learning process between teacher and student, and make higher education an even less happy and warm environment. It would make students and staff strangers to each other. It's absolutely disgusting that anyone would be contemplating their usage."*

In summarising this sub-theme, there was a tendency throughout the responses to assume that deepfakes would be used by institutions maliciously and with cost-saving in mind, leading to an eventual ousting of the academic and an automation of higher education to maximise profitability on behalf of the institutions, while at the same time creating divisions between larger and smaller institutions. This was summarised by one respondent as follows:

*"It doesn't seem like it is offering us the opportunity to do something that we ought to be doing in the first place; just a cheaper way to do higher education badly. I say cheaper, but I presume access to the technology will be prohibitively expensive; so big institutions will be able to afford it, and therefore run a leaner, more profitable operation, and consolidate their grip in the higher education "marketplace", at the expense of students and teachers."*

*Sub-theme: Societal Impacts*

Aside from the devaluation of the educational experience and the denigration of the role of the academic, many responses expressed concern over larger societal issues caused by deepfakes. Often, this related to the energy used to generate AI outputs:

> "Not to mention the environmental impact of the massive energy drain used in creating them (72 hours for 30 seconds of a somewhat-convincing Anderson Cooper!)."

> "The enormous environmental impact of using energy-intensive AI technologies to generate content that likely already exists online or could be more economically delivered via standard methods."

However, this was not the only societal impact mentioned in the responses. An overall decline in the validity of information and the ability to distinguish real from false was a prominent topic, and other responses also drew on the potential for exacerbating digital divides across institutions.

> "If students become accustomed to deepfake technology in the classroom, they might struggle to distinguish between real and fabricated content outside of it. In addition, the use of deepfake technology might exacerbate the technical and financial divide between education providers."

**Theme 3 – Modulating Educator Efficiency**

*Sub-theme: Workload*

While the majority of responses reflected serious ethical concerns regarding the use of deepfakes, with many arguing that no potential benefits could outweigh the significant and far-reaching consequences of their use in education, some responses noted the potential for alleviating academic burdens:

> "Deepfakes could be used to tweak or update existing videos with little effort or to quickly make videos with improved sound and visual quality if we have access to software and training."

> "If the AI is used to make the delivery neater and tidier -- or more accessible (e.g. in a different language) -- then that seems OK."

These examples suggest that there may be a perception of a modest benefit if such technologies are used to effectively automate existing tasks when creating multimedia content.

*Sub-theme: Cost-benefit*

However, the cost-benefit of deepfake technology was often depicted negatively. Given that deepfakes are rapidly developing and the time and energy required to create them are decreasing, this position may change in the future. However, a common pattern throughout the responses was that time would be better spent elsewhere to improve the educational experience for learners:

> *"Time investment...specifically the return on investment. Is a deepfake asset in my class worth the time and effort it took to make it (learning gains, class satisfaction, etc.)"*

**Qualitive Analysis Summary**

The thematic analysis constructed three overarching themes that describe the tensions surrounding deepfake technology in higher education. The first of these reflects that respondents feel that deepfakes may have the potential to benefit inclusion and student engagement, yet these are moderated by the overall degrading effects that could occur with misuse, such as exploitation and devaluation of the academic role, and the automation of higher education leading to job loss and poorer provision for students, along with larger planetary and societal impacts. The third theme also exemplified this contradictory set of ideas, with some responses highlighting the potential for liberation from administrative tasks (e.g. video editing or delivery of content), while others questioned the price-value of the investment in new technologies. The relationships between these themes reveal a fundamental tension between educational innovation and institutional responsibilities. Stakeholders weighed the potential pedagogical benefits against significant concerns about power dynamics, ethical implementation, and professional autonomy, suggesting the need for a carefully considered approach to deepfake adoption in higher education.

## Discussion and Integration of Findings

Our study showed complex and sometimes contradictory attitudes towards deepfake technology in HE. The quantitative findings showed that adoption intentions were primarily driven by hedonic motivation, with a gender-moderated effect on price value sensitivity. Qualitative data revealed significant concerns regarding ethical implications and institutional power dynamics. This tension between enjoyment and apprehension offers important insights into how emerging AI technologies, such as synthetic media and deepfakes, might be integrated into higher education.

**Hedonic Motivation and Technology Acceptance**

The emergence of hedonic motivation as the strongest predictor of deepfake adoption intention aligns with studies on educational technology adoption (Deng & Yu, 2023; Granić, 2022; Nikolopoulou et al., 2021) that highlight the importance of enjoyment in technology acceptance, particularly for novel technologies such as deepfakes. While deepfakes are often associated with harmful applications such as misinformation and harassment (Burkell & Gosse, 2019; Chesney & Citron, 2019; Harris, 2021), our findings suggest that educators can distinguish between malicious uses and potentially beneficial educational applications. This may reflect a selection bias in our sample; educators willing to participate in technology surveys may be more inclined to technological experimentation. However, this concern is somewhat tempered by the number of responses that explicitly opposed the use of the technology. The enjoyment factor might also stem from educators recognition of deepfakes' potential to reduce workload while creating engaging content, as evidenced by our qualitative findings on efficiency and workload reduction, echoing the views of Westerlund (2019). This mirrors findings from research on other novel technologies, where perceived enjoyment can outweigh concerns about perceived usefulness (Holdack et al., 2022).

**Professional Identity**

A particularly interesting tension emerged between hedonic motivation and concerns about 'diminishing human value' in education. While educators expressed enjoyment in experimenting with deepfake technology and recognised its potential benefits in efficiency, they worried about its potential to automate and devalue the teaching profession. As one participant noted, deepfakes would *"devalue the personalised learning process between teacher and student and make higher education an even less happy and warm environment."* This speaks to broader anxieties about the automation of academic work, with participants expressing fears that institutions would exploit deepfake technology to use their content or identities without employing them to do so. This reflects the complex professional identity of educators, who are expected to engage with new technologies to support students while balancing innovation with their own perceived authentic practice. This suggests that adoption decisions are influenced not only by traditional technology acceptance factors, but also by educators' fears and doubts about the potential harm of the teaching profession brought about by these novel technologies.

**Resource Inequity and the Digital Divide**

Another clear ethical dimension that emerged from our analysis was resource inequity and institutional access to deepfake technology. While GenAI technologies have been proposed as a way to potentially democratise inequities in education (Gesser-Edelsburg et al., 2024; James & Andrews, 2024; Perkins, 2023; Roe, 2024), our findings suggest that for deepfakes specifically, there is a more complex reality. The current landscape of GenAI technology, characterised by high computational requirements and significant licencing costs, raises critical questions regarding educational equity and access (James & Andrews, 2024; Perkins et al., 2024) and the implications of these resource disparities extend beyond immediate access. As deepfake technology becomes more sophisticated and potentially more integral to educational content creation, institutions that cannot invest in these tools may find themselves at a competitive disadvantage in terms of both student recruitment and retention.

**Theoretical Implications and Framework Development**

While the regression model identified only hedonic motivation and price value as significant predictors of intention to adopt deepfake technology, this reflects the robustness of the UTAUT2 framework in explaining deepfake adoption. As the model explained 75% of the variance in educators' adoption intentions, this indicates strong predictive power. The analysis also revealed moderate relationships between most UTAUT2 constructs and adoption intentions, with the non-significance of some predictors in the regression model stemming from shared variance rather than a lack of influence. For example, while hedonic motivation emerged as particularly influential (uniquely explaining 24% of the variance), the remaining variables collectively explained 51% of the variance in adoption intentions, even without this predictor.

However, the quantitative model alone does not fully capture the nuanced ethical considerations that emerged from our qualitative analysis. While UTAUT2 effectively explains what drives adoption intentions, it does not address the deeper concerns about professional identity, institutional power dynamics, and ethical implications that educators expressed. This suggests that while existing technology acceptance frameworks remain valuable for understanding adoption decisions, they may need to be complemented by new approaches that explicitly incorporate ethical dimensions when studying emerging AI technologies, such as deepfakes.

**Deepfake Adoption Framework**

Any implementation of deepfake technologies in HEIs must be driven by clearly identified pedagogical needs or benefits, rather than technological capability alone. Before deployment, institutions should conduct thorough pilot studies and engage in comprehensive discussions with relevant governance committees. The significant concerns demonstrated in our research indicate that the decision to implement deepfake or other synthetic media technologies should not be taken lightly. If these technologies are to be considered for implementation, we recommend a structured approach in three key areas:

First, robust institutional policies and governance structures must be established prior to implementation. Given educators' significant concerns about potential misuse, institutions must develop clear policies regarding consent, transparency, and the ethical use of deepfakes in educational contexts. These policies should explicitly address fears regarding job security by delineating appropriate use cases and establishing safeguards against replacing human educators. Regular assessments should monitor the impact on educational quality and professional relationships, with clear procedures for addressing the misuse of this technology. Policies should emphasise identifying clear educational benefit use cases before widespread deployment, ensuring meaningful pedagogical value beyond mere novelty.

Second, professional development and support systems are required for the sustained adoption of these methods. Our finding that hedonic motivation strongly predicts adoption intentions suggests that initial enthusiasm may not translate into long-term meaningful use without proper support. Professional development should focus on pedagogical integration and ethical considerations alongside technical competence, and support systems should help educators progress beyond initial experimentation to develop applications that genuinely make a difference for educators and students. Given our findings on gender differences in technology acceptance, these programmes should incorporate inclusive approaches that address diverse needs and concerns.

Third, resource allocation must prioritise equity and evidence-based implementation. Although deepfake technologies currently require significant investment, costs are likely to decrease as the technology matures and open-source alternatives emerge. Rather than rushing to widespread adoption, institutions should conduct pilot studies to evaluate the educational impact and identify use cases that specifically support underserved students, and educators who might gain the most value from these tools. Implementation decisions should be guided by evidence of improved learning outcomes rather than technological capability alone. Investment should prioritise applications that demonstrably reduce existing technological divides rather than exacerbate them.

This is summarised in Figure 1.

**Figure 1: Deepfake Adoption Framework**

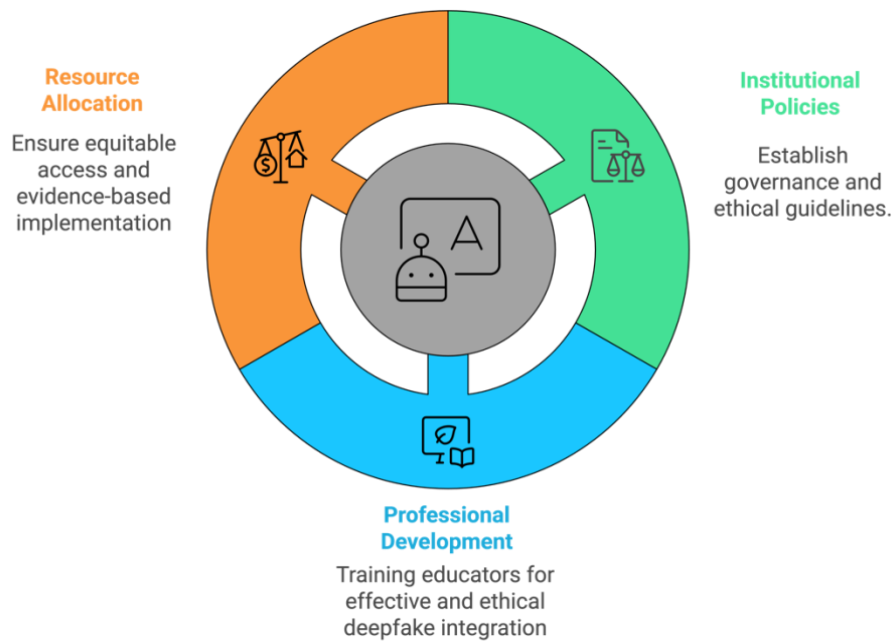

**Limitations and Future Research Directions**

Several methodological limitations of our sample characteristics require careful consideration in future studies. The relatively modest sample size ($n = 174$) after data cleaning, while adequate for our approach, may have limited our ability to detect smaller effects, particularly in interaction analyses. More critically, the self-selection nature of our sampling strategy likely introduced systematic bias into our findings. Given that the survey explicitly focused on deepfake technology in education, respondents may have been motivated to participate based on particularly strong positive or negative views of this technology, potentially skewing our results towards more polarised perspectives. This self-selection bias may be especially relevant given the controversial and current nature of deepfake technology and its ethical implications for the workplace. Additionally, our recruitment methods, primarily through professional networks and institutional email lists, may have excluded educators who were less engaged with the chosen networks. The timing of our data collection, which was conducted during a period of intense public discourse on AI in education, may have also influenced response patterns, particularly regarding concerns about automation and job security. These sampling limitations suggest the need for future research employing more diverse recruitment strategies and perhaps embedding deepfake-related questions within broader surveys about educational technology to capture more representative viewpoints.

To address these limitations and advance our understanding of this domain, future research should pursue the following key directions. First, scholars should develop new theoretical models specifically designed to capture the ethical complexities of emerging educational technologies, moving beyond traditional acceptance frameworks to incorporate institutional power dynamics and cultural variables into their research. Second, longitudinal investigations are needed to track the evolution of attitudes and adoption patterns as deepfake technology matures and becomes more prevalent in educational settings. Third, research must be expanded to include students' perspectives and experiences, providing insights into the impact of deepfake technology on learning outcomes and educational engagement.

# Conclusions

This study provides important insights into an underexplored yet rapidly developing area of AI in the context of higher education. Our quantitative findings highlight the primacy of hedonic motivation and gender-specific cost considerations in adoption intentions, and our qualitative analysis reveals deeper concerns about professional identity and resource inequity, echoing broader debates about AI tools in education. These findings suggest that traditional technology acceptance models may require substantive revision to adequately capture the complexity of emerging AI technologies in education. Our results highlight the need to consider ethical dimensions alongside traditional acceptance factors. The emergence of only two significant predictors from the UTAUT2 model, combined with insights into institutional power dynamics and resource inequities, suggests that future theoretical frameworks must move beyond individual-level technology acceptance to consider broader systemic, policy, and structural factors. As policies around the ethical use of Gen AI applications are still nascent (Bjelobaba et al., 2024; Perkins & Roe, 2023, 2024), having frameworks in place can support institutions and educators in dealing with these new and unknown technologies.

Furthermore, our findings demonstrate the paradoxical nature of technological innovation in the higher education sector. While deepfake technology shows promise for enhancing educational experiences through increased engagement and potentially reduced workload, its adoption is complicated by multiple intersecting tensions: enjoyment and ethical concerns, professional identity and innovation, and individual and institutional interests. These tensions reflect the broader challenges in the ongoing digital transformation of higher education, where technological capabilities must be balanced against pedagogical integrity, institutional equity, and professional autonomy. The implications of this study extend beyond the immediate context of deepfake technology. As educational institutions increasingly grapple with emerging AI technologies, our findings suggest a framework for deepfake adoption that explicitly addresses issues of resource allocation, institutional policy, and professional development. Future development and implementation of these tools should balance these competing factors while prioritising educational integrity and human relationships in teaching and learning.

Finally, although we maintain the importance of distinct studies on deepfake technology and higher education stakeholders, we also assert that there is significant work to be done in this area and in the area of AI in education more broadly. To fully understand the ramifications of deepfake technology, efforts must be made to explore students' voices in potentially using this technology to engage with a programme of study in the future. Although there is a growing understanding of student perspectives on GenAI tools (Albayati, 2024; Roe, Perkins, & Ruelle, 2024; Shoufan, 2023), we must expand this research to examine student views specifically regarding deepfake and synthetic media technologies, along with AI literacy (Kumar et al., 2024; Ng et al., 2021; Roe et al., 2025) to ensure minimisation of potential impacts from deepfake misuse in higher education settings.

***Declaration of Generative AI and AI-assisted technologies in the writing process***

GenAI tools were used for ideation and in some passages of draft text creation which was then heavily revised, along with editing and revision during the production of the manuscript. The tools used were ChatGPT (o3-mini-high) and Claude 3.5 Sonnet, which were chosen for their ability to provide sophisticated feedback on textual outputs. Napkin.ai was used to design Figure 1. These tools were selected and used supportively and not to replace core author responsibilities and activities. The authors reviewed, edited, and take responsibility for all outputs of the tools used.

# Appendix 1

**Table 4: Item Wording and UTUAT2 Construct Assessed**

| Construct | Item |
|---|---|
| Performance Expectancy | Management at my university have indicated that they want educators to use deepfakes in their teaching practices in the near future. |
| Performance Expectancy | I believe deepfakes could be used effectively to create engaging educational content. |
| Effort Expectancy | Learning to use deepfake technology in an education context would be easy for me. |
| Social Influence | Colleagues whose opinions I value would support the use of deepfakes in education. |
| Facilitating Conditions | I have the resources necessary to use deepfakes in education. |
| Facilitating Conditions | I have the knowledge necessary to use deepfakes in education. |
| Hedonic Motivation | I would enjoy creating or using deepfake content for educational purposes. |
| Price Value | I believe that incorporating deepfakes into my teaching will take more time than it's worth.* |
| Habit | I typically make use of new/innovative technology in my teaching. |
| Behavioural Intention | I would use deepfakes as part of my teaching practice if the option became available to me. |

* Item reverse coded